\documentclass[9pt,arxiv]{lapreprint}
\pdfoutput=1

\usepackage[version=4]{mhchem} 
\usepackage{siunitx}    
\usepackage{pdflscape}  
\usepackage{rotating}   
\usepackage{textgreek}  
\usepackage{gensymb}    
\usepackage[misc]{ifsym} 
\usepackage{orcidlink}  
\usepackage{listings}   
\usepackage{colortbl}   
\usepackage{tabularx}   
\usepackage{longtable}  
\usepackage{subcaption}
\usepackage{multirow}
\usepackage{snotez}     
\usepackage{csquotes}   

\DeclareSIUnit\Molar{M}

\usepackage[			
	backend=biber,      
    style=ieee,   
	natbib=true,		
	hyperref=true,	    
	alldates=year,      
]{biblatex}

\title{Toward Models of Impact and Recovery of the US Western Grid from Earthquake Events}

\author[ \orcidlink{0000-0002-8158-9317} 1]{Riley Weinmann}
\author[ \orcidlink{0000-0002-3964-3260} 1]{Eduardo Cotilla-Sanchez}
\author[ \orcidlink{0000-0002-3093-108X} 1 \Letter]{Ted K.A. Brekken}

\affil[1]{School of Electrical Engineering \& Computer Science, Oregon State University}

\correspondence{ted.brekken@oregonstate.edu}{Ted Brekken}

\presentaddress[]{School of Electrical Engineering \& Computer Science, Oregon State University}

\funding[]{This research was funded by the National Science Foundation, award number 1933217.}
\compint[]{The author declare no competing interests.}

\leadauthor{Weinmann}
\shorttitle{Toward Models of Impact and Recovery of the US Western Grid from Earthquake Events}

\begin{document}
\maketitle

\begin{abstract}

A Cascadia Subduction Zone (CSZ) earthquake will cause widespread damage to numerous lifelines and infrastructure along the northern US west coast.
The goal of the presented research is to provide a \textit{bottom up} estimate of the impact on and subsequent recovery of a Cascadia Subduction Zone earthquake on the US western grid to supplement and enhance the expert opinion estimates provided to date.
The scope is limited to consideration of shaking damage to utility substation equipment components of a power system model.
The analysis utilizes probabilistic models of damage and recovery for substation power system assets, along with graph techniques for modeling connectivity, and Monte Carlo quasi steady state power flow solutions.
The results show that a conservative estimate of the initial damage and loss of load is approximately 4,000 MW, with a recovery estimate of 230 days.

\end{abstract}

\section{Introduction}

\begin{figure}
    \centering
    \includegraphics[height=0.40\textheight,width=0.95\textwidth,keepaspectratio]{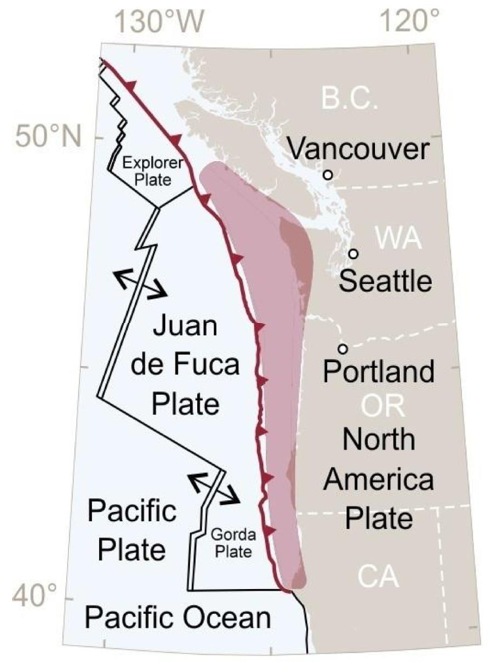}
    \caption{The Cascadia Subduction Zone, which has the capacity to release a +9M earthquake.  (Image from \cite{crew}.)}
    \label{fig:crew_csz}
\end{figure}

The Cascadia Subduction Zone (CSZ) -- shown in Figure~\ref{fig:crew_csz} -- is an approximately 1,000~km long offshore fault, stretching from Cape Mendocino in northern California through the states of Oregon and Washington to the Brooks Peninsula in southern British Columbia \cite{orp,exercise,crew,goldfinger}. 
The fault is part of a ring of of subduction zones that surrounds the Pacific Ocean, creating a formation called the ``Ring of Fire."
The fault bears many similarities to its geological mirror, the large subduction fault east of Japan that released the +9M Tohuku earthquake and tsunami in 2011, which killed nearly 20,000 people, cost several 10s of billions of USD in damage, and left approximately 4 million homes without power in the immediate aftermath \cite{Kazama:2012aa}.

It is estimated -- based on geological observations -- the last CSZ earthquake and tsunami occurred around 1700 AD \cite{orp}.
Considering the geological record further back to 8000 BC, the mean time between CSZ earthquakes is around 300 years, but with a fair degree of variability, in which some earthquake events may be separated by 100 years, and some by more than 500 years.
The calculated odds that the next earthquake will occur in the next 50 years range from 7\% to 15\% for a ``great" earthquake affecting the entire PNW to about 37\% for a ``very large" earthquake affecting southern Oregon and northern California \cite{orp}.

What is certain is that a CSZ earthquake and accompanying tsunami will damage or destroy many lifeline systems along the coast:~transportation, communication, water, sewer, etc, and the electrical system.
A CSZ magnitude 9 earthquake would induce several minutes of shaking along nearly the entire northern and central west coast, with surface Peak Ground Accelerations (PGA) reaching 1 g and higher -- much higher in some localized cases.
The severe shaking will induce highly damaging stresses in electrical bushings and connections (i.e., circuit breakers, switches, and transformers), equipment support structures, and substation bus-work, and can damage auxiliary generation systems.
The shaking will also induce liquefaction and landslides that threaten transmission towers and distribution poles.

In 2011, the Oregon legislature passed House Resolution 3 to understand the full extent of the impact that this hazard can pose to the state and devise a plan to make the state more resilient.
This effort resulted in the Oregon Resilience Plan (ORP) \cite{orp}.
The ORP estimates that electrical services could be out for up to 6 months on the coast after a CSZ earthquake.
This estimate is provided by expert opinion.

This paper aims to provide a more comprehensive estimate of the electrical impacts of a CSA earthquake, and subsequent recovery, using a detailed Monte Carlo damage study of the entire US western grid.
The goal is to provide a \textit{bottom up} estimate of the impact on and subsequent recovery of a Cascadia Subduction Zone earthquake on the US western grid to supplement and enhance the expert opinion estimates provided to date.

\section{Background}

There are extensive resources on modeling earthquake and other natural disaster impacts to structures and systems, including electrical infrastructure \cite{Pires:1995aa,S.A.Zareei:2017aa,Oikonomou:2016aa,Huo:1995:Seismic-Fragility,Ersoy:2008aa,hazus,Rojahn:1991:Seismic-Vulnerability,Hines_2008,wang_2016}.
IEEE Standard 693 focuses purely on requirements and certification of seismic hazard compliant substation equipment \cite{ieee-693-2018}.
Additionally, there are numerous reconnaissance reports on earthquake impacts to electricity systems, particularly through databases such as the Multidisciplinary Center for Earthquake Engineering Research (MCEER) at the University of Buffalo \cite{Kwasinski:2014aa,Park:2006:Nisqually-Earthquake,Kazama:2012aa,Asfura:1999:The-Quindio-Columbia,Goltz:1994:The-Northridge-California,Schiff:1998:The-Loma-Prieta,Lee:2000:The-Chi-Chi-Taiwan,Cimellaro:2013:Emilia-Earthquake}.  
Much of this work focuses on the impact to an individual component of a substation or aggregates expected component damage to estimate  damage to a substation.  
However, research that connects electrical system structural modeling to power system model connectivity and power flow solutions is limited.

The methodology used in this paper is similar to that used in \cite{PMAPS_2022} where the power system bus-branch model resolution is improved with the addition of electrical bays. 
This paper uses multiple substation archetypes, adding redundancy at each bus. 
Additionally this paper uses fragility functions and restoration functions created for specific substation components e.g., circuit breakers. 
This further improves the model and allows for failure and restoration evaluation with improved resolution.  
Both methods improve on the standard bus-branch (BB) model by analyzing the connectivity of the substations represented by each bus.

\section{Western US Grid Model}

\begin{figure}
    \centering
    \includegraphics[height=0.40\textheight,width=0.95\textwidth,keepaspectratio]{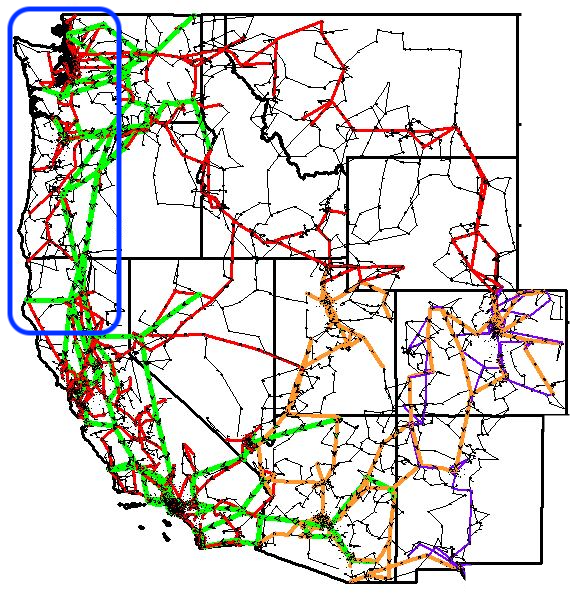}
    \caption{The geographical layout of the ACTIVSg10k synthetic test case.  The blue rectangle denotes the Damage Zone: the area most strongly impacted by a Cascadia Subduction Zone earthquake. Image adapted from \cite{Birchfield2018}.}
    \label{fig:ACTIVSg10k}
\end{figure}

The ACTIVSg10k test case is a synthetic bus-branch model of the US western grid interconnect \cite{Birchfield2017}.
A diagram of the model is shown in Figure~\ref{fig:ACTIVSg10k}. 
The model is an open source synthetic model and is used to enable open dissemination of research methodologies and results.

The ACTIVSg10k case has 10,000 buses, 12,706 branches and 2,485 generators. 
Generation data comes from the U.S.~Energy Information Administration. 
The 10,000 buses are mapped to 4,762 geographic locations in the Western United States, and the load for each bus is estimated from US government census data.   

\begin{table} 
\caption{ACTIVSg10k system summary\label{tab1}}
\newcolumntype{C}{>{\centering\arraybackslash}X}
\begin{tabularx}{\textwidth}{CCCC}
\toprule
 & \textbf{Buses}	& \textbf{Gen.~Capacity [MW]}	& \textbf{Load [MW]}\\
\midrule
\textbf{ACTIVSg10k}	    & 10,000	& 209,420      & 150,917\\
\textbf{Damage Zone}	    & 1,459		& 20,140       & 23,122  \\

\bottomrule
\end{tabularx}
\end{table}

The impact of a CSZ earthquake will be focused on a sub-region of the model in the the Pacific Northwest region of the United States, defined as the region from the coast to the $121^{st}$ meridian between the $49^{th}$ and $39^{th}$ parallels, which is approximately the rectangle bounded by the US-Canada border in the north, Bend OR in the west, and Mendocino County CA in the south.
This area will be referred to as the \textit{Damage Zone}.
In the ACTIVSg10k model the Damage Zone region includes 728 geographic locations for a total of 1,459 buses, which is approximately 15\% of the baseline model.

\section{Fragility, Restoration, and Metrics}

To properly use the ACTIVSg10k model for this study, each of the thousands of assets in the model requires an associated fragility and restoration function to provide the likelihood of damage and subsequent recovery.
This section details the functions and data sources used to model seismic fragility and subsequent restoration, as well as the metrics used to quantify the system damage and recovery.

\subsection{Fragility Functions}\label{FailureFunctions}

\begin{figure}
    \centering
    \includegraphics[height=0.40\textheight,width=0.95\textwidth,keepaspectratio]{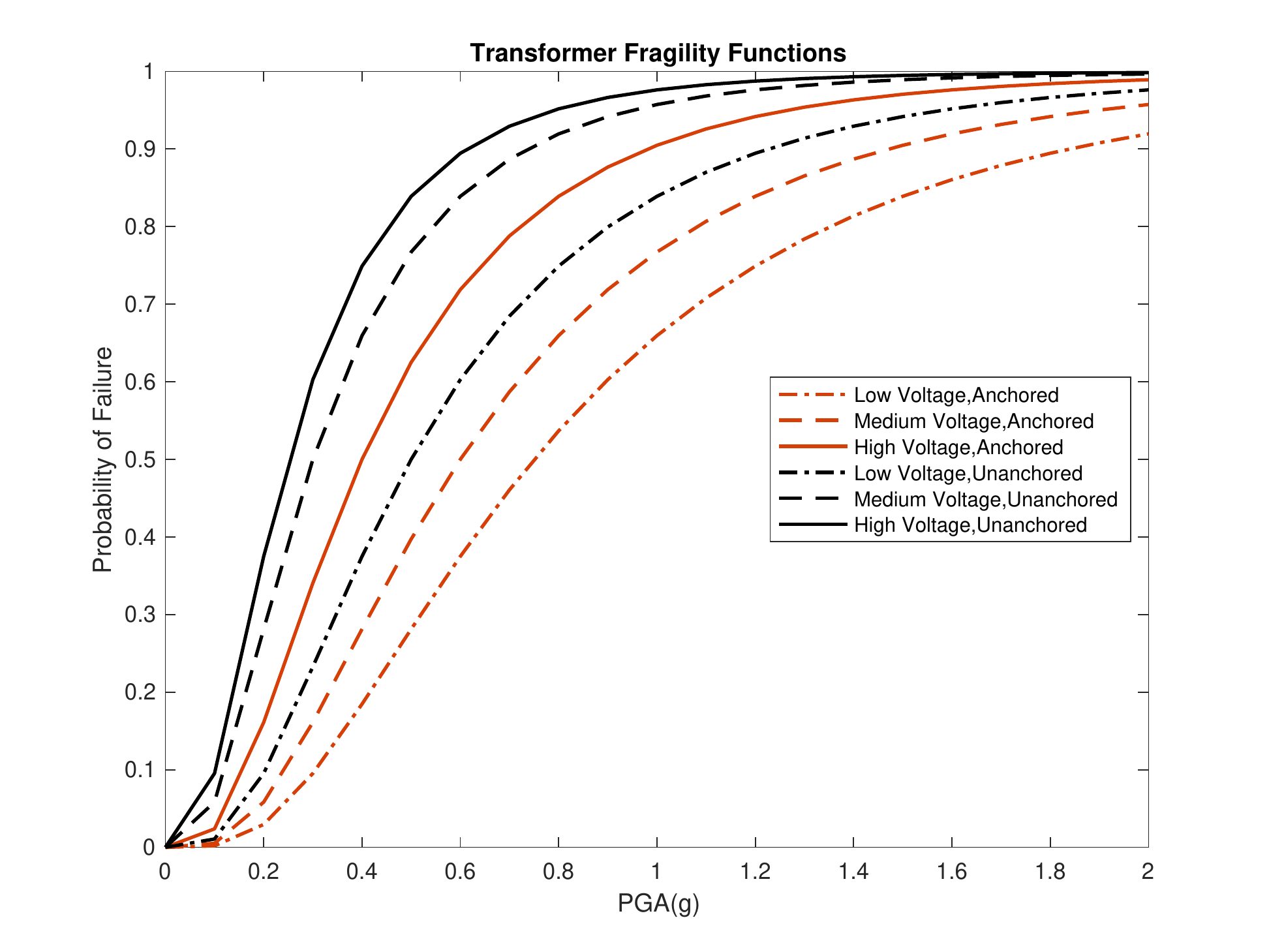}
    \caption{Transformer fragility functions, plotted with parameters from Hazus \cite{hazus}.}
    \label{fig:ff_transformer}
\end{figure}

Fragility functions represent the probability that a component will be in failed or damaged state after experiencing a certain level of exogenous excitation, which is usually expressed as Peak Ground Acceleration (PGA).
In this research, we use power system component fragility functions from Hazus \cite{hazus}, which are classified for transformers, circuit breakers, and disconnect switches for three voltage levels:~34.5 kV to 150 kV (low voltage); 150 kV to 350 kV (medium voltage); and 350 kV to 765 kV (high voltage).
The low voltage level is extended down to 13.2 kV, the lowest voltage in the ACTIVSg10k model.
An example fragility function is shown in Figure~\ref{fig:ff_transformer}.

Transmission and distribution towers and poles were not included in the analysis, as historically those components generally perform well with regards to shaking stress and damage.
(The greater threat to towers and poles is not shaking, but landslides and liquefaction \cite{Lee:2000:The-Chi-Chi-Taiwan,Kwasinski:2014aa}, which are not modeled here. 
Please see \nameref{sec_assumptions} for further discussion.)

To assess the damage status of a component, first the PGA experienced by component at that location is determined from a PGA map for a specific seismic event.
The PGA maps are traditionally determined by civil and geotechnical engineers by modeling the amount of energetic release from an earthquake fault and subsequent propagation of the energy through the bedrock and overlying soil.

The M9.0 Cascadia Subduction Zone scenario shakemaps produced by Wirth et al. \cite{Wirth:2020:Ensemble-ShakeMaps} were utilized for estimates of peak ground acceleration for each bus and substation for the regional analysis. 
Rather than consider only one rupture scenario, these estimates of ground motions were derived from a logic tree approach considering multiple earthquake rupture scenarios, which were produced by Frankel et al.~\cite{Frankel:2018:Broadband-synthetic}, in order to account for uncertainty.  
Wirth et al.~applied adjustments for site response based on the soil and geologic conditions based on mapped Vs30 estimates produced by Heath et al.~\cite{Heath:2020:A-global-hybrid}. 
Amongst a variety of products produced, the US Geological Survey provides rasters of both the log of the mean PGA and the uncertainty (log of standard deviation) spatially mapped across the western US. 
Distant locations outside of the raster boundaries that are still part of the electric grid network were assumed to experience no significant shaking given that the PGA values converge to 0 g at the edges of the raster. 

\begin{figure}
    \centering
    \includegraphics[height=0.40\textheight,width=0.95\textwidth,keepaspectratio]{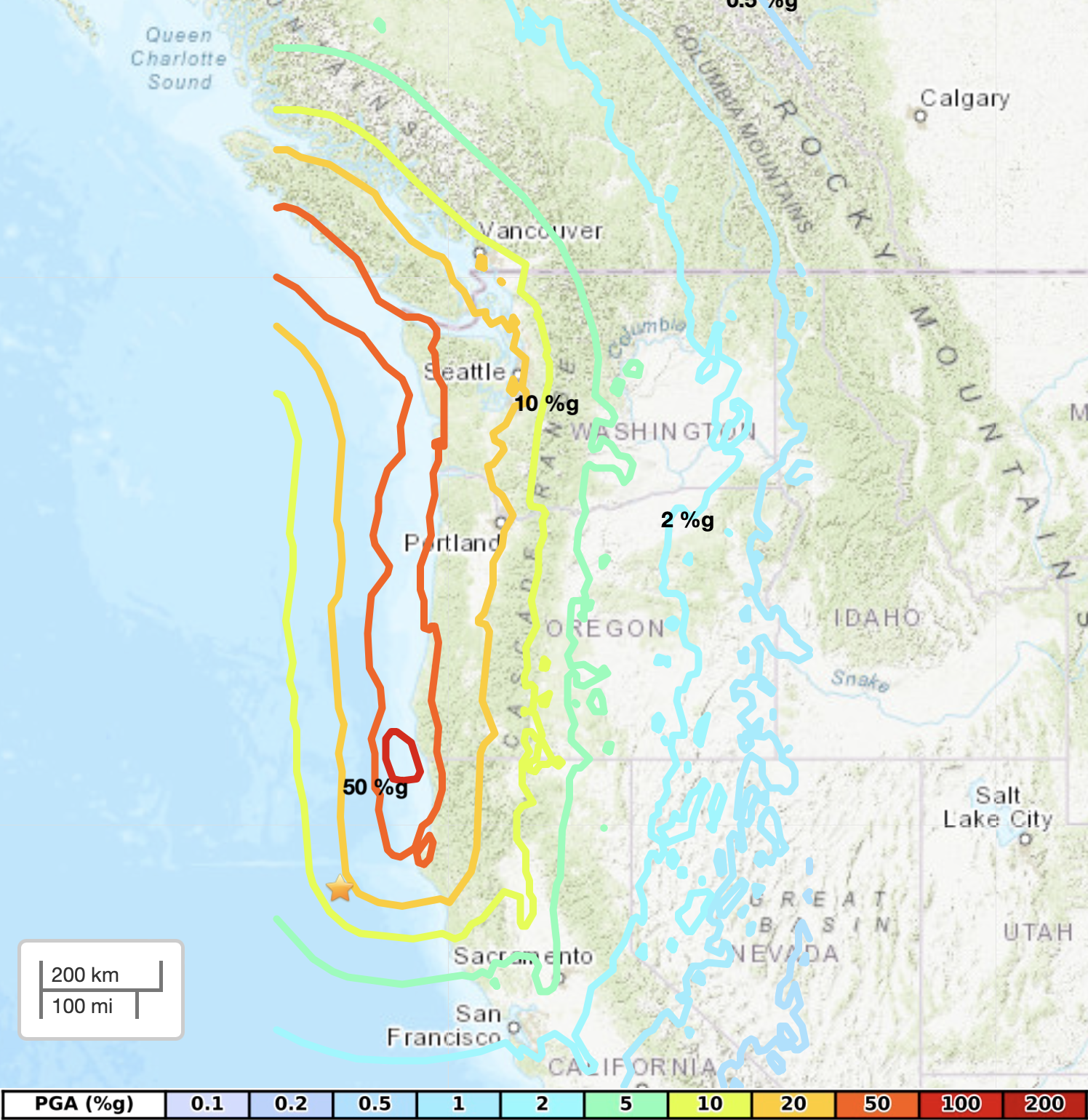}
    \caption{PGA map for a magnitude 9 CSZ event. Image from the US Geological Survey (USGS) Earthquake Hazards Program website.  The PGA values in the legend along the bottom of the figure are in \% of 1 g.}
    \label{fig:usgsM9}
\end{figure}

As an example for visualization, the PGA map for the M9.0 Cascadia Subduction Zone scenario is shown in Figure~\ref{fig:usgsM9}.

\subsection{Restoration Functions}\label{RestorationFunctions}

As part of this research, we conducted a survey of electric utility industry experts in the United States Pacific Northwest region to determine approximate restoration time tables for each component at various voltage levels. 
Survey respondents provided a time frame for restoration or replacement of severely damaged or destroyed components. 
Using these survey responses and a log-normal fit, we generate curves describing the probability that an asset would be recovered on a given day post-earthquake.
Respondents were asked to assume that necessary personnel were available and each site was accessible, therefore considerations of road or workforce inaccessibility, damage to access roads, and damage to other lifelines (e.g., communications systems) is not factored into the analysis.
\textit{Therefore the restoration results from the current version of the proposed model based on these survey results should be considered as a best-case scenario.
It is very likely that cross-lifeline failures and other unmodeled factors will cause additional delays to repairs beyond the estimates used in this analysis.}
Because these additional delays and interactions are very difficult or impossible to estimate at this point in time, the analysis will proceed without them. 
Future planned work will increase the resolution of the model by considering not only delays and infrastructure interactions but also the lifecycle of the components. 

\begin{figure}
    \centering
    \includegraphics[height=0.40\textheight,width=0.95\textwidth,keepaspectratio]{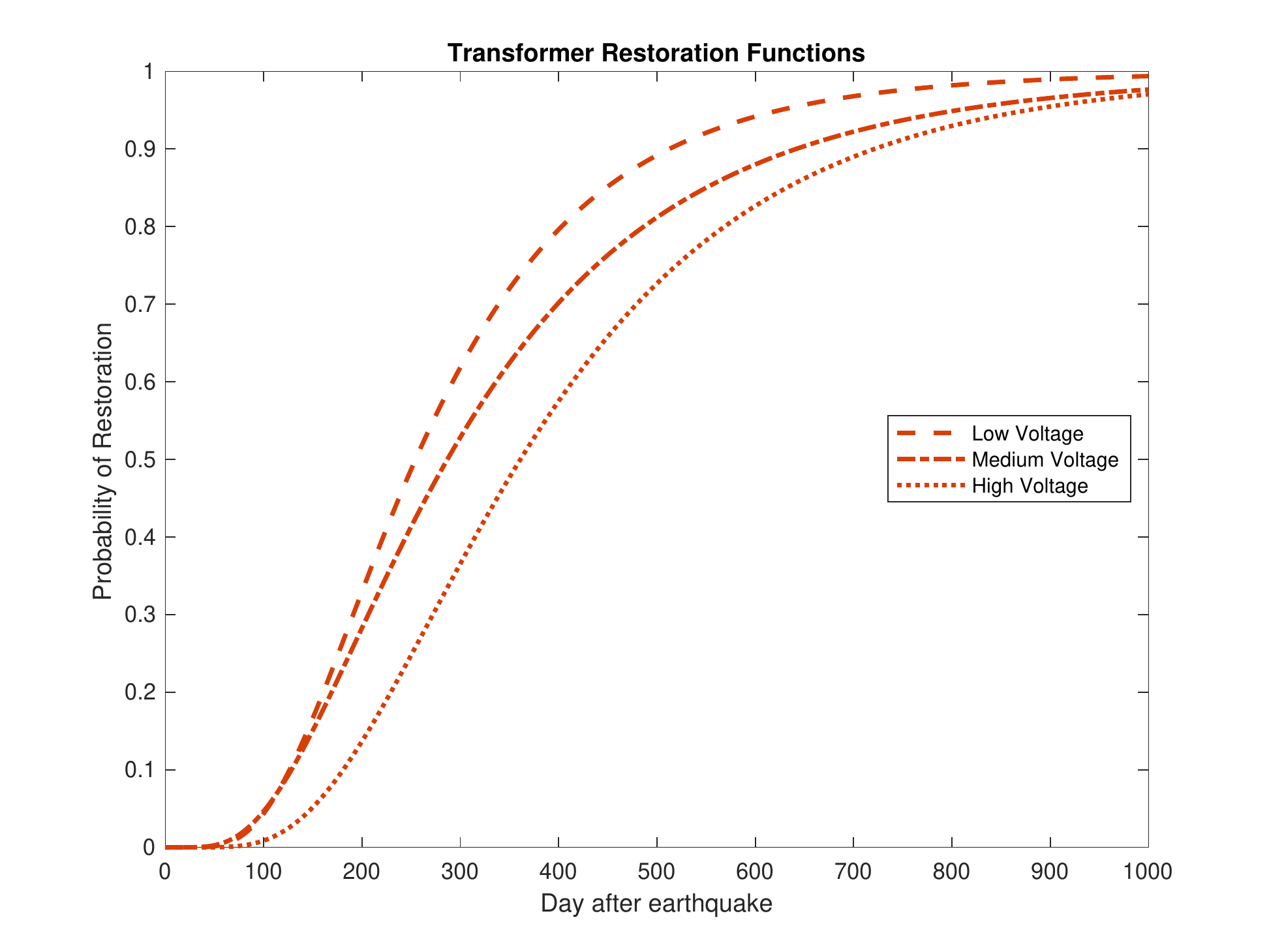}
    \caption{Transformer restoration functions, created with log-normal fit of survey data.}
    \label{fig:examplerestfunc}
\end{figure}

An example restoration function is shown in Figure~\ref{fig:examplerestfunc}.
Restoration functions are used to probabilistically establish the recovery day after the earthquake event for each failed component.
After determining that an asset is in a damaged or failed state via the asset fragility function, a random number is drawn from a uniform distribution and is associated with the cumulative probability on the y-axis, which then indicates the number of days until restoration via the corresponding value on the x-axis.
As an example using Figure~\ref{fig:examplerestfunc}, a uniform random number sample of 0.60 would indicate a given high-voltage transformer will be recovered on approximately day 430 after the earthquake.
The recovery times of all damaged assets are determined at the start of the iteration.

\subsection{Resilience Metrics} \label{sec:resiliencemetrics}

\begin{figure}
    \centering
    \includegraphics[height=0.40\textheight,width=0.95\textwidth,keepaspectratio]{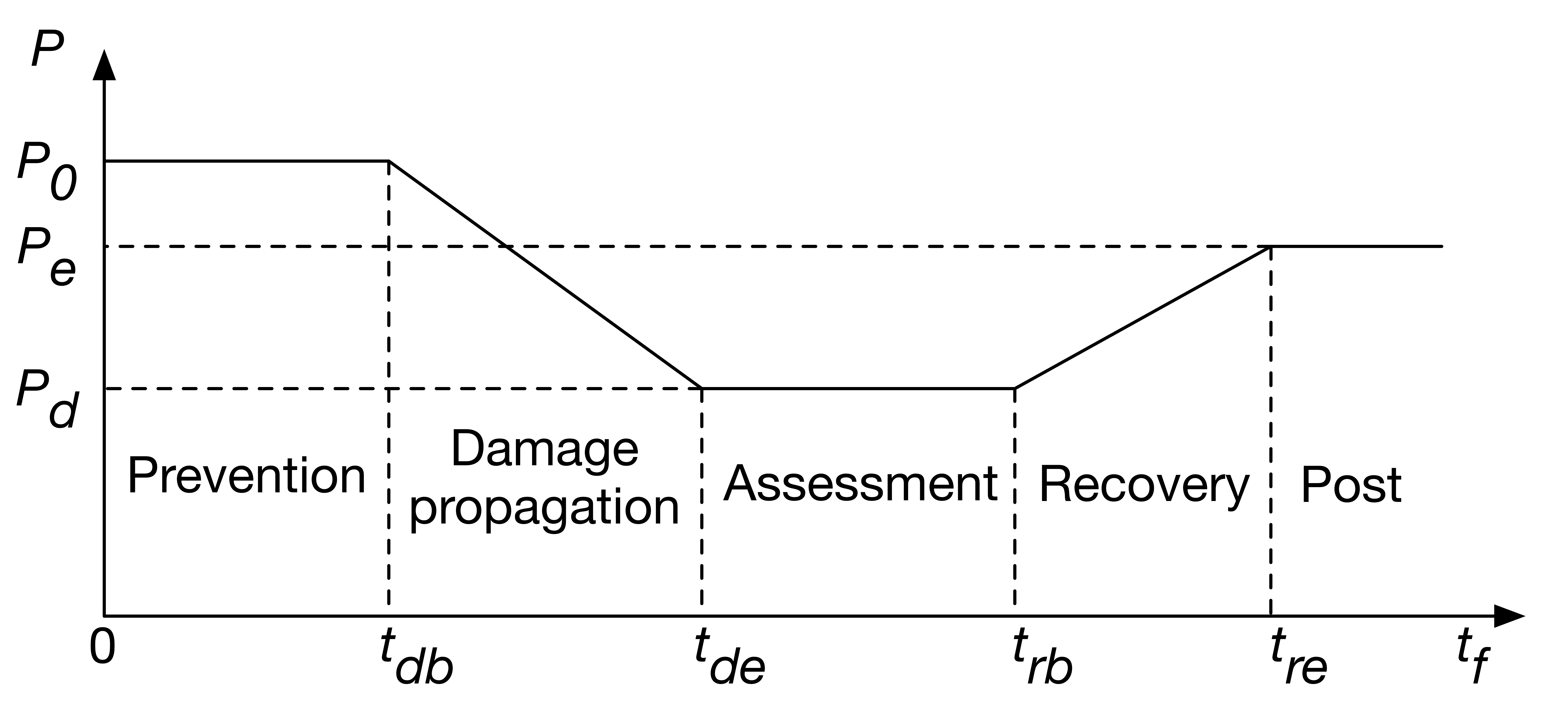}
    \caption{An example system response to a damaging event at time $t_{db}$.  The variable $P$ is some performance quantity of the system.  In a power system case, it could typically be power delivered to load.  (Image adapted from \cite{Panteli:2017aa,Panteli:2017ab}.)}
    \label{fig_performance_curve}
\end{figure}

A core set of resilience quantifiers \cite{Panteli:2017aa,Panteli:2017ab} can be derived from the function of the system performance vs.~time, as shown in Figure~\ref{fig_performance_curve}.

The variable $P$ represents some measure of system performance.
In the case of a power system, a typical choice would be amount of power successfully delivered to system load over time.
A damaging event occurs at time $t_{db}$ and causes damage to the system until time $t_{de}$.
There is then a period of assessment in which resources are collected and restoration actions and strategies are planned.
Restoration then takes place over the time $t_{rb}$ to $t_{re}$, at which point the system enters a new steady state, likely with a different level of performance than before.

From this performance trajectory, several resilience metrics can be derived.
\begin{equation}
    \Phi = \frac{P_d - P_0}{t_{de} - t_{db}} \quad \mbox{(rate of performance decrease)}
\end{equation}
\begin{equation}
    \Lambda = P_0 - P_d \quad \mbox{(amplitude of performance decrease)}
\end{equation}
\begin{equation}
    \mathcal{E} = t_{rb} - t_{de} \quad \mbox{(time in damage state)}
\end{equation}
\begin{equation}
    \Pi = \frac{P_e - P_d}{t_{re} - t_{rb}} \quad \mbox{(rate of performance increase)}
\end{equation}
\begin{equation} \label{eq:resilienceR}
    R = \frac{\int_0^{t_f} P(t) dt}{P_0 \cdot t_f} \quad \mbox{(resilience)}
\end{equation}

\section{Analysis Method}

\begin{figure}
    \centering
    \includegraphics[height=0.85\textheight]{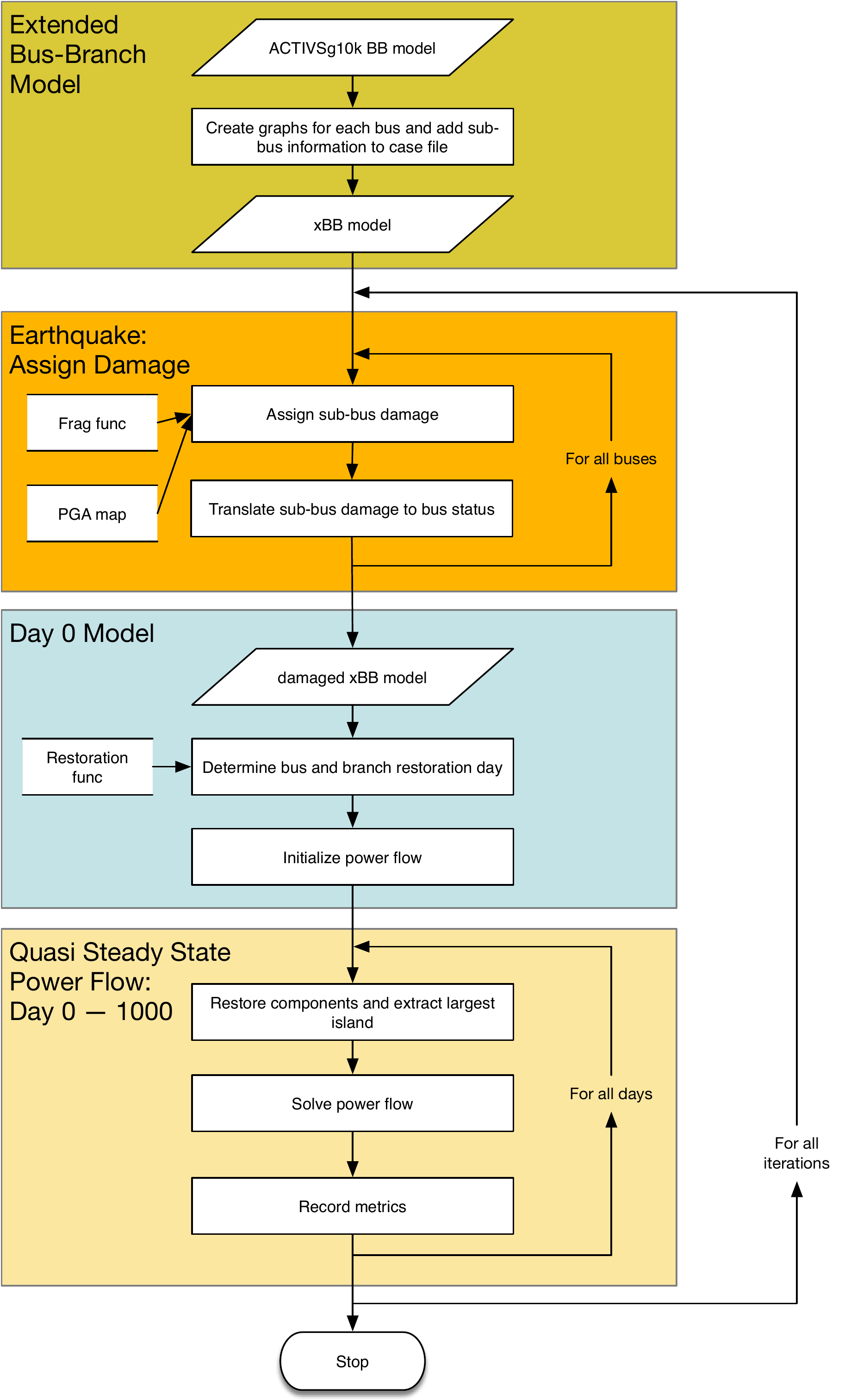}
    \caption{Overall analysis and simulation structure. Each iteration is a quasi-steady state solution over 1,000 days, for 2,500 iterations total.}
    \label{fig:flowchart}
\end{figure}

Damage and recovery from a CSZ earthquake was modeled using a Quasi Steady-State (QSS) Monte Carlo methodology. Figure \ref{fig:flowchart} shows a flow diagram of the simulation architecture. 

Prior to beginning the simulation, the bus-branch model resolution is improved by an extended bus-branch (xBB) model that models each bus as a graph of connected components.
See \nameref{sec:xBB} for details. 
For each iteration of the simulation, first damage is inflicted on the components of the xBB model using component fragility functions. 
Then, the damage to the xBB model is translated to the bus-branch model for power flow analysis. 
Immediately following damage phase the system begins recovery. 
For each damaged component in the xBB model a restoration function determines the day that component will be recovered.  
The simulation then steps through each day of restoration. After each simulated day of restoration, the total connected load and status of system assets are logged.
Additionally, the largest island is extracted from the BB model and a power flow solution is calculated and logged, if convergent.

\subsection{Extended Bus Branch Model}\label{sec:xBB}

To improve on the resolution of the BB model we create an extended bus-branch (xBB) model by associating each bus in the ACTIVSg10k test case with a graph.
Graph connectivity is determined according to the four substation archetypes described in \cite{chalishazar_augmenting_2018} and the number of feeders connected to each bus. 
The xBB model, which includes a graph for each bus, provides substation component level resolution allowing for partial functionality of buses, which would not be possible in a traditional BB model.  

For the purpose of creating the substation graph, we count feeders as any BB asset connected to the bus. 
Buses with two feeders are considered Single Bus Single Breaker (SBSB), buses with three or four feeders are modeled as a Ring Bus (RB), buses with five or more feeders are modeled as Breaker and a Half (BAH). 
Any bus connected to a generator is modeled as Double Bus Double Breaker (DBDB). 
For example, a bus that in the BB model has two generators, one load, and two branches would have five feeders and be modeled as a DBDB configuration due to the presence of generation at the bus.

\begin{figure}
    \centering
    \includegraphics[width=0.52\textwidth]{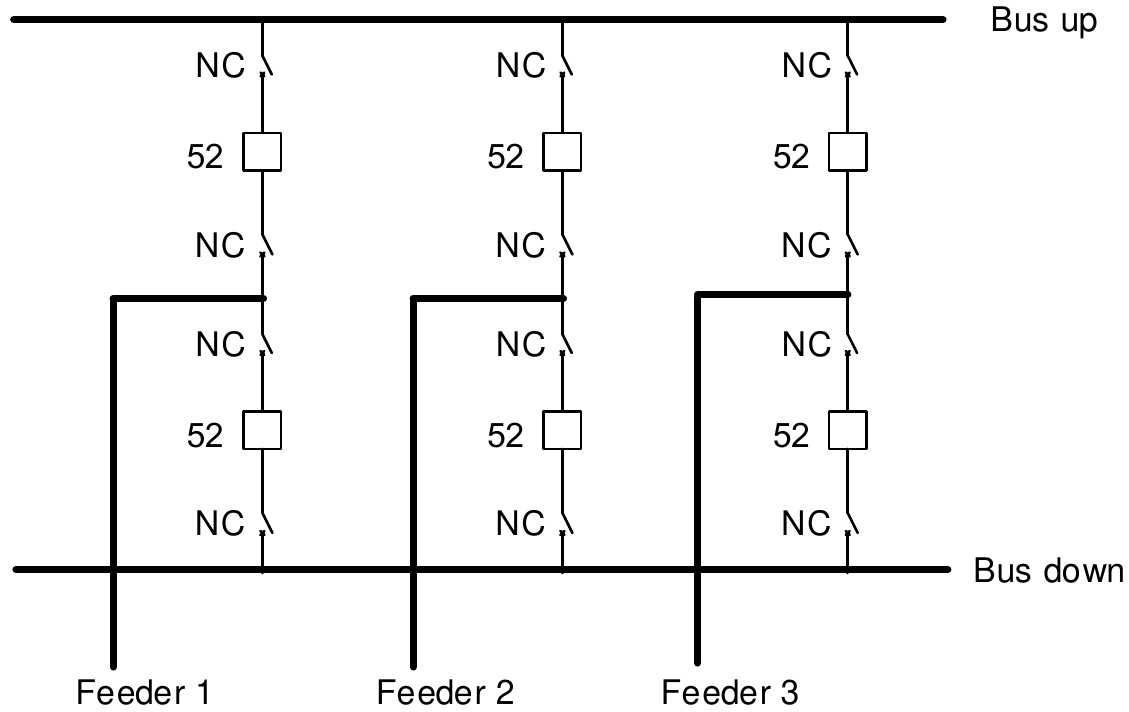}
    \hfill
    \includegraphics[width=0.44\textwidth]{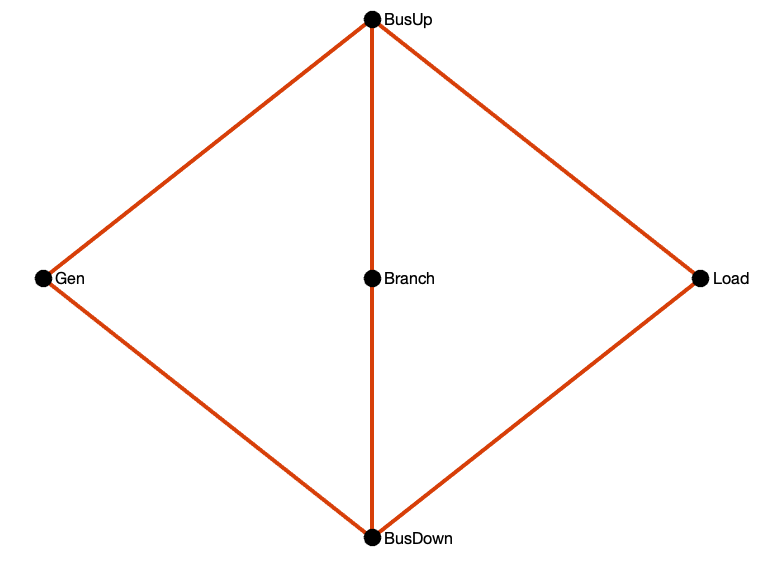}
    \caption{Example graph representation of a double-bus double-breaker substation, with feeder 1 connected to a generator, feeder 2 connected to another substation, and feeder 3 connected to a load.}
    \label{fig:xBBgraph}
\end{figure}

In a given graph, nodes represent bus-branch (BB) level assets (e.g., generator, load, branch) and edges represent the series connected components connecting each asset according the substation archetype.  
Connections between BB assets are assumed to be the series connection of a two switched disconnects (SD) and a circuit breaker (CB). 
This SD-CB-SD grouping is considered a switch group (SG) which is modeled by an edge in the xBB substation graph. 
Figure~\ref{fig:xBBgraph} shows an example of a substation graph with 5 nodes and 6 edges. 
This graph models a DBDB configured substation with 3 feeders, connected by 12 SDs and 6 CBs.  

\subsection{Damage Implementation}

Damage is implemented in two stages. 
First, components of the xBB model are damaged using PGAs sampled from a log-normal distribution of a M9.0 CSZ earthquake PGAs and asset failure functions as discussed in \nameref{FailureFunctions}.
Component failure results in the removal of corresponding edge in the substation graph. 
Second, the connectivity of each substation graph is evaluated and nodes with a degree of zero have the corresponding BB asset removed from service.

\subsection{Damage Recovery}

Recovery begins immediately after the CSZ event and continues for 1,000 days. 
Components that were not damaged during the damage state are considered to be recovered on day zero. 
At each time step components with a day of restoration occurring on that day are recovered. 
Once all of the components in a switch-group (SG) have been recovered (on that day or previously) the corresponding edge is returned to the substation graph. 
At the end of each day the connectivity of each substation graph is evaluated. 
If the degree of a node is no longer zero, the asset corresponding to that node is returned to service. 
A substation is said to be functionally connected once the degree of all nodes is greater than zero. 
This indicates that a viable connection exists between all assets in the BB model. 
For a substation to return to full service, all damaged components must be recovered.

\subsection{Power Flow Analysis}

At the end of each recovery day the number of assets out of service and the total load connected to the system is logged.
In addition, an attempt is made to find a power flow solution using MATPOWER \cite{MATPOWER}.
To improve the probability of finding a converging solution, the power flow is calculated for the largest island in the system. 
This island is assumed to be the interconnected portion of the WECC system still intact following a CSZ earthquake. 
As recovery progresses eventually only one island will remain containing all 10,000 buses. 
If a solution converges, the load served is calculated by summing power demanded at each bus and  generation available calculated as the sum of the maximum generation at each bus.

\subsection{Assumptions and Limitations} \label{sec_assumptions}

Modeling and predicting the damage and subsequent restoration of electrical system components in a large earthquake is exceedingly complicated, and dependent on many correlated factors.
To enable a first-order approximate analysis, several simplifications were made:
\begin{itemize}
    \item The analysis only considers shaking damage to electrical substation components of the power system model.
    \item Damage to the load is not modeled, only its connection to the grid through substation equipment.
    Therefore the eventual recovery is to the full pre-earthquake load.
    \item Transmission and distribution towers and poles were not included in the analysis, as historically those components generally perform well with regards to shaking stress and damage.  The greater threat to towers and poles is liquefaction and landslides, which are not modeled here.
    \item Restoration times did not explicitly include damage to roads and transportation lifelines in the same area, or the impacts of damage to communication and fuel systems.  (I.e, no interdependent lifeline failures were modeled.)
    \item Restoration times assumed qualified technicians and engineers were available immediately after the earthquake.  (I.e., there was no delay to the beginning of the restoration process and therefore the \textit{assessment} time of Figure~\ref{fig_performance_curve} is 0 days.)
    \item Equipment lead times (and therefore the restoration timelines) are determined from industry expert survey results which are based on typical non-emergency lead-times.  In the case of an earthquake, lead-times may be impacted by the large simultaneous demand for many types of the same equipment, as well as emergency supplies provided from utilities across the country.
    \item Federal emergency response, which may accelerate repair and restoration, was not modeled.
\end{itemize}

All considered, it is therefore very likely that the results presented in this paper represent an optimistic scenario.
The authors consider that the expected extent of damage, and the restoration time, would likely exceed the median scenarios presented here, and thus the authors recommend consideration of the 95th percentile results as a conservative estimate of the likely damage and recovery.

\section{Results}

\subsection{Earthquake Impact and Recovery Overview}

Significant shaking is confined to the region extending east from the coast to the $121^{st}$ meridian between the $49^{th}$ and $39^{th}$ parallels, which is approximately the rectangle bounded by the US-Canada border in the north, Bend OR in the west, and Mendocino County CA in the south. Outside of this region shaking is expected to be slight to none  -- and therefore no significant damage -- beyond this region. 
This affected sub-area of the model will be referred to as the \textit{Damage Zone}, and some of the results in this section will be presented with respect to this sub-area of the WECC.
Note that, because the ACTIVSg10k model is a synthetic model, the bus locations shown do not necessarily correspond to actual physical substations.
Instead, the results shown should be interpreted as indicative of trends over a wide geographical area of similar expected extent as the actual WECC grid.

\begin{figure}
    \centering
    \includegraphics[height=0.40\textheight,width=0.95\textwidth,keepaspectratio]{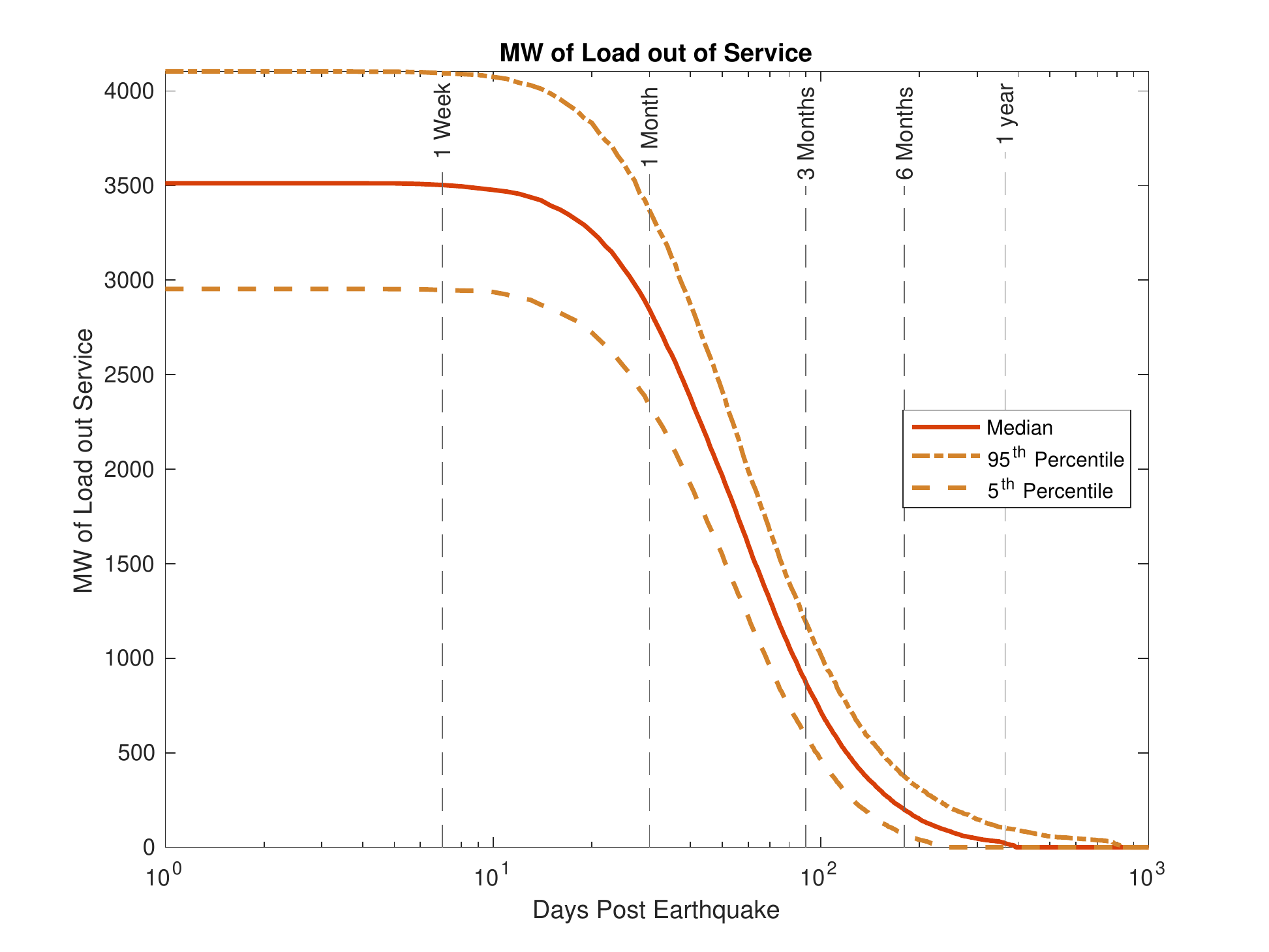}
    \caption{Restoration of load vs time, including the 5th and 95th percentile cases.}
    \label{fig:load_restoration}
\end{figure}

\begin{figure}
    \centering
    \includegraphics[height=0.40\textheight,width=0.95\textwidth,keepaspectratio]{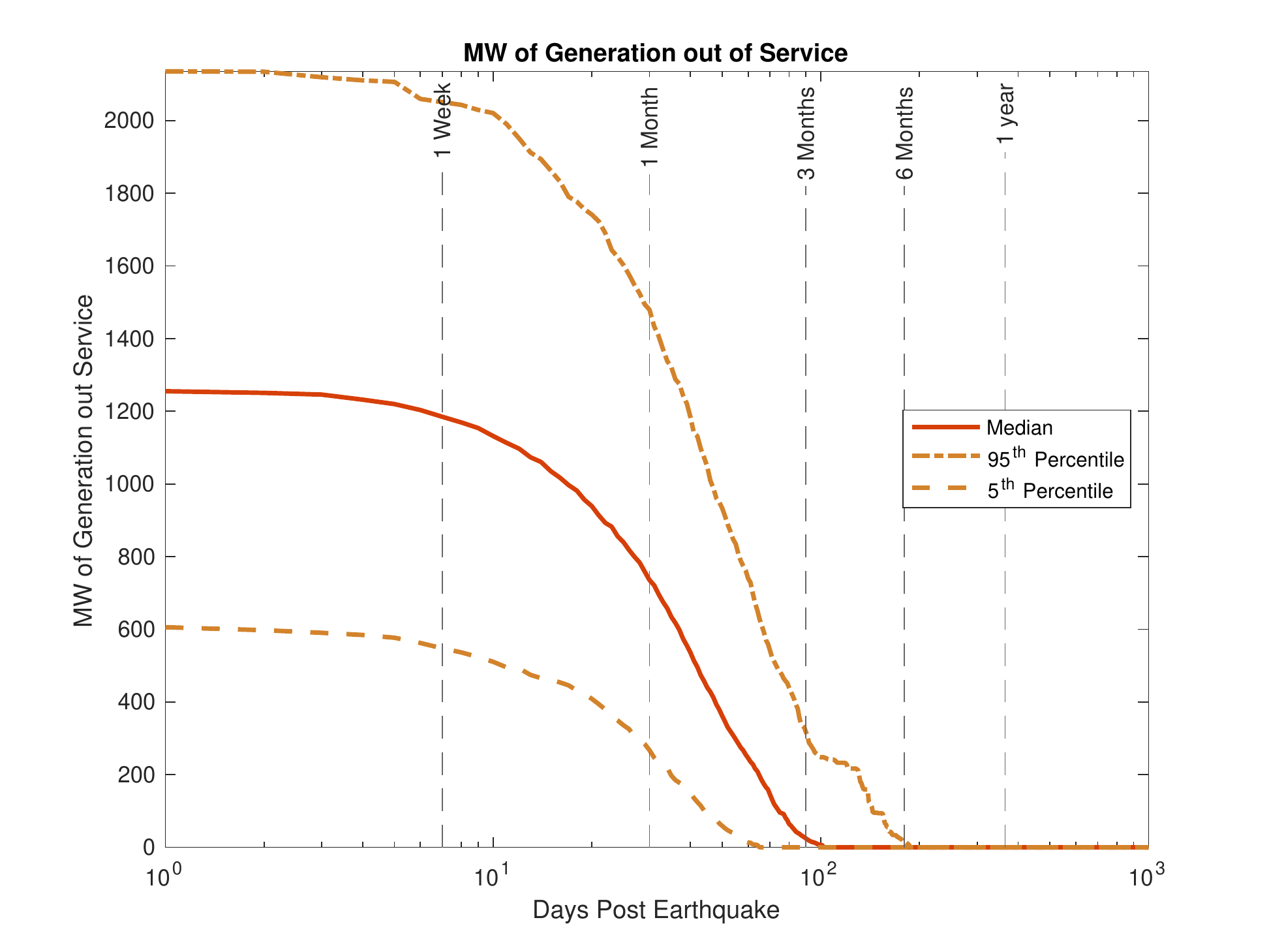}
    \caption{Restoration of generation vs time, including the 5th and 95th percentile cases.}
    \label{fig:gen_restoration}
\end{figure}

The initial earthquake impact in terms of load and generation lost, and subsequent recovery, is shown in Figure~\ref{fig:load_restoration} and \ref{fig:gen_restoration}.

It can be seen that the most significant rate of recovery occurs between 1 and 3 months, with a large majority of recovery by 6 months, and nearly complete recovery within a year.

\begin{figure}
    \centering
    \includegraphics[height=0.50\textheight,width=0.95\textwidth,keepaspectratio]{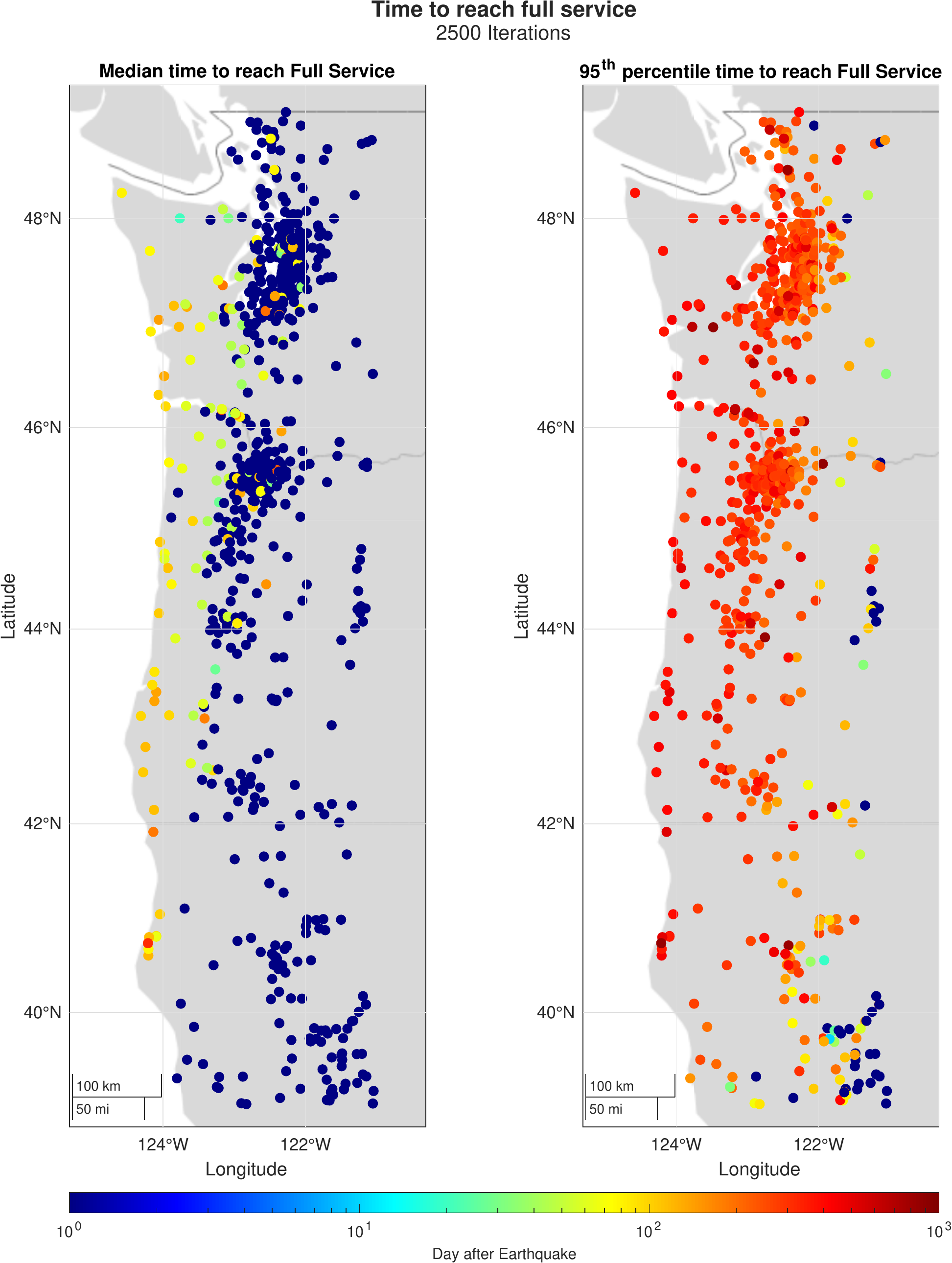}
    \caption{The median and 95th percentile time for each substation in the synthetic model to reach full service (i.e., the full restoration of every asset in the substation).}
    \label{fig:heatmap}
\end{figure}

The recovery can be further visualized in Figure~\ref{fig:heatmap}, which shows the median and $95^{th}$ percentile full restoration time of substations in the ACTIVSg10k model.

Generally, it can be observed that substations located closer to the coast will have longer recovery times compared to those located further inland. 
It is shown that the median time to recover full substation functionality (i.e., service) is approximately 100 to 200 days, and the 95th percentile time to recover full functionality is approximately 6 months.
It is noted that time to reach a lower function of service, in which the minimum breakers and switches required for any level of substation function are replaced or repaired, will be less.

\subsection{Analysis Metrics}

\begin{figure}
    \centering
    \includegraphics[height=0.40\textheight,width=0.95\textwidth,keepaspectratio]{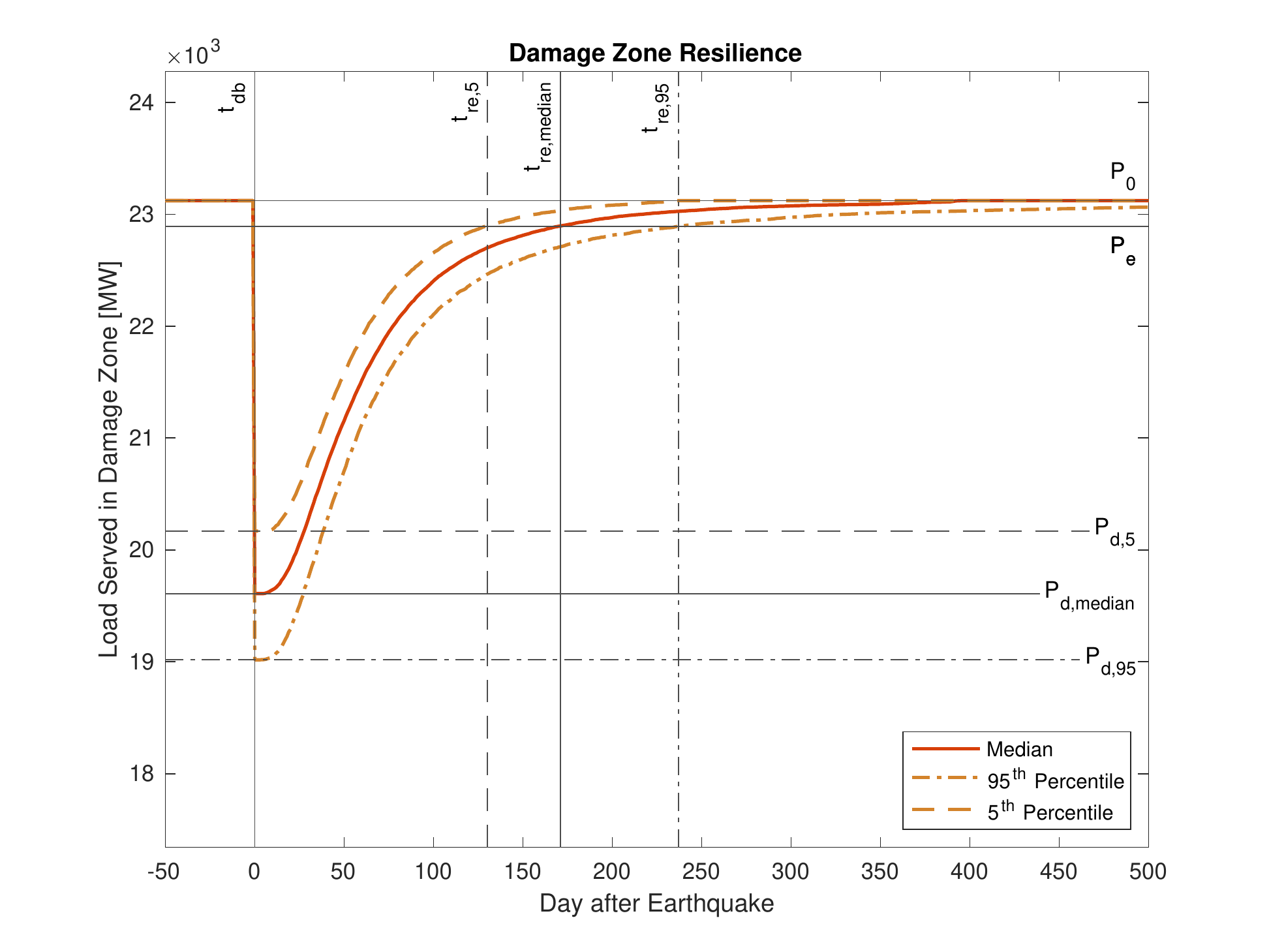}
    \caption{Load still connected to the main system within the Damage Zone.}
    \label{fig:resilience}
\end{figure}

In this section we deploy the metrics defined in \nameref{sec:resiliencemetrics}.
The performance metric $P(t)$, as shown in Figure~\ref{fig_performance_curve}, is defined as load connected in the \textit{Damage Zone} sub-area of the system, normalized to the undamaged load in the same zone.
The performance metric result, including the median and 5th and 95th percentile results, is shown in Figure~\ref{fig:resilience}. 

From the curve, we calculate the metrics defined in \nameref{sec:resiliencemetrics}.
Because the load recovers asymptotically, we have selected a recovery to 99\% of the pre-earthquake load as the criteria for which the system is recovered: $P_e = 0.99 \cdot P_0$.
For calculating and comparing the resilience metric $R$, the final time $t_f$ is set to the recovery end time of the 95$^{th}$ percentile.

\begin{table} 
\caption{Resilience metrics}
\begin{tabularx}{\textwidth}{llccc}
\toprule
Metric & Metric description & 5\%	& 50\% (median)	& 95\% \\
\midrule
$P_0$ & Initial performance [MW] & 23,122 & 23,122 & 23,122 \\
$P_d$ & Damaged performance [MW]  & 20,169 & 19,609 & 19,018 \\
$P_e$ & Recovered performance [MW] &22,891 &22,891 &22,891 \\
$t_{db}$ & Damage start time [day] &0 &0  &0 \\
$t_{re}$ & Recovery end time [day] &130 &171 &237 \\
$\Lambda$ & Performance decrease [MW] & 2,953 &3,512 & 4,104  \\
$\Pi$ & Rate of recovery [MW/day] & 20.9 & 19.2 & 16.3 \\
$R$ & Resilience &0.962 &0.951 &0.937 \\
\bottomrule
\end{tabularx}
\end{table}

\subsection{Power Flow Results}

The power flow results are more difficult to obtain as the heavily damaged system shortly after the earthquake is often non-convergent.
Therefore the analysis proceeds as above, with gradual recovery of the system until it reaches a state with a stable solution.

\begin{figure}
    \centering
    \includegraphics[height=0.40\textheight,width=0.95\textwidth,keepaspectratio]{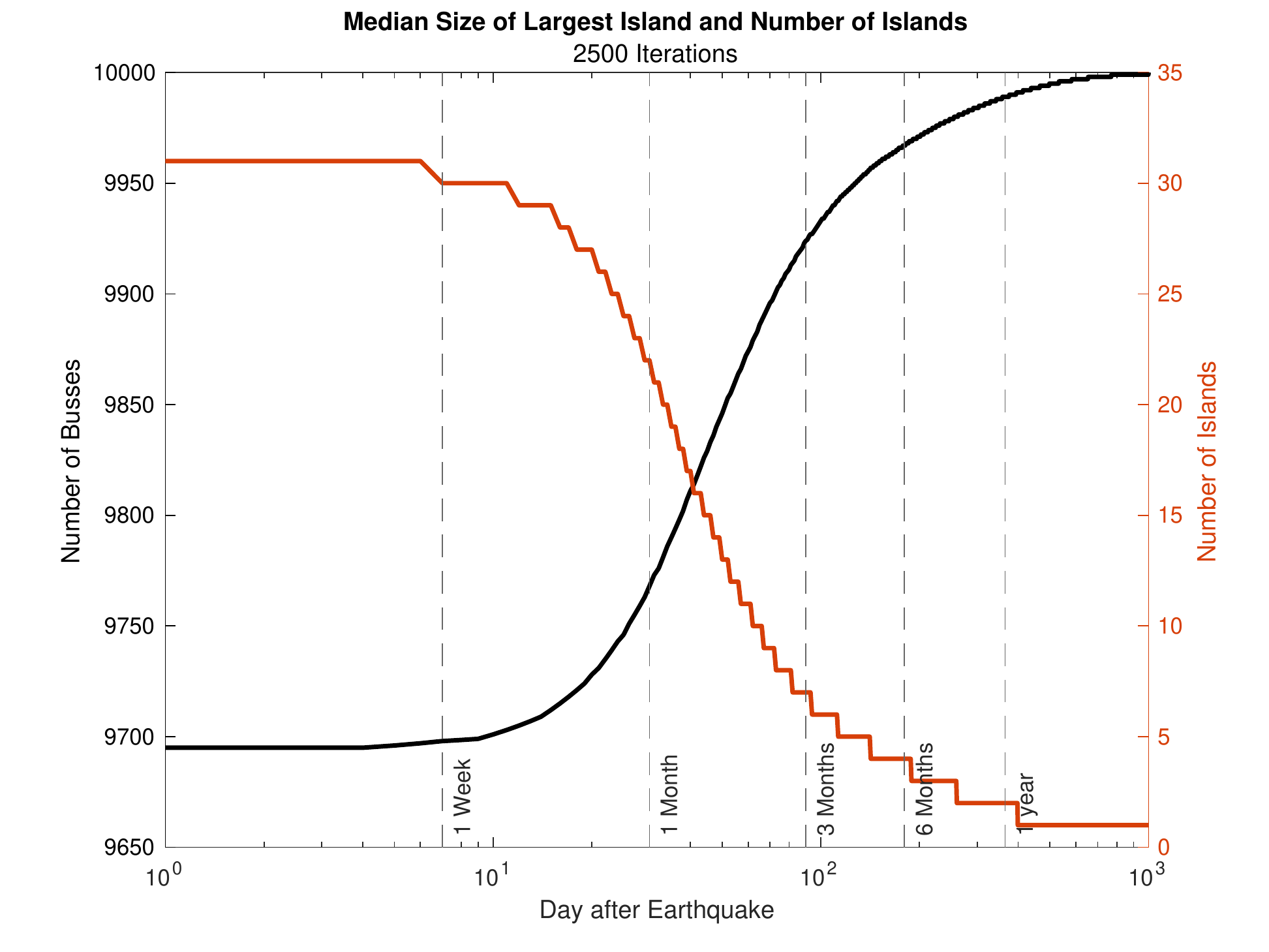}
    \caption{Number of islands, and size of the largest island for the entire western grid model.}
    \label{fig:islands}
\end{figure}

\begin{figure}
    \centering
    \includegraphics[height=0.40\textheight,width=0.95\textwidth,keepaspectratio]{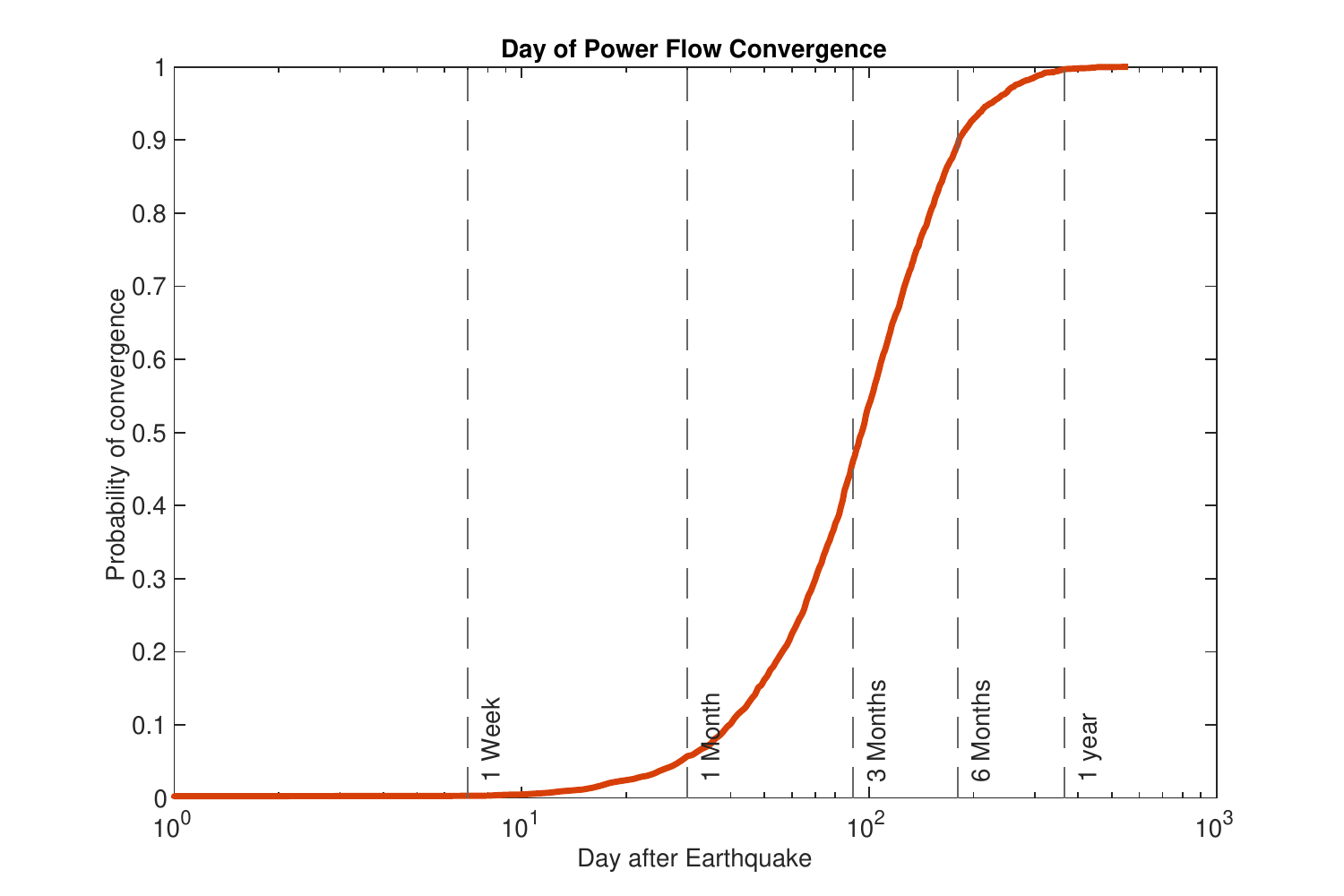}
    \caption{Cumulative distribution function of power flow convergence over the 1,000 day restoration period for 2,500 iterations.}
    \label{fig:powerflowconvergence}
\end{figure}

The Figures~\ref{fig:islands} and \ref{fig:powerflowconvergence} show the median size of the largest island, as reported by MATPOWER, as well as the median convergent power flow load served at the soonest day a convergent result is available.

The results show that initially a convergent power flow solution is not attainable.
When a convergent solution is available, it generally verifies the delivery of load equal to the load connected to the main island as shown in Figure~\ref{fig:resilience}.

This illustrates the limitations in using a regular power flow solver to assess system status, as even light damage to a large complex power flow case can cause non-convergence.
Thus, additional tools and metrics are required to quantify system status.

\subsection{Summary and Analysis}

This paper presents a methodology for estimating the impact and subsequent recovery of the US western grid to a Cascadia Subduction Zone earthquake.
The methodology uses Monte Carlo analysis to account for the probability of shaking, damage, and recovery of each component of the system from the bottom up.
The methodology also utilizes a novel graph-based sub-methodology to determine the functionality of substations.

The results show that overall damage to the affected sub-area of the western grid (i.e., the Damage Zone) is estimated to cause a $95^{th}$ percentile approximate loss of 4,000 MW immediately after the earthquake, with a subsequent recovery of approximately 230 days.
The area of damage is largely limited to the sub-region of the model from the coast to the $121^{st}$ meridian between the $49^{th}$ and $39^{th}$ parallels, which is approximately the rectangle bounded by the US-Canada border in the north, Bend OR in the west, and Mendocino County CA in the south.

In terms of power flow, the system is initially split into approximately 30 islands -- one large island and many small islands -- and is gradually recovered to one system again in approximately 1 year.
The largest island is often not convergent until nearly 3 months or more of recovery, thus illustrating a high degree of sensitivity in solving power flows on large, complex power system models in a heavily damaged state.
This illustrates the need for islanded power flow system metrics and quantifiable measures beyond power flows for heavily damaged systems.

It is important to note the various assumptions made in this study, as discussed in \nameref{sec_assumptions}.

The authors wish to thank the following for their support and advisement in this research: Mike Olsen, Armin Stuedlein, and Eve Lathrop of Oregon State University; Jay Landstrom of Portland General Electric; Shamus Gamache of Central Lincoln PUD; Leon Kempner and David Glennon of the Bonneville Power Administration; and Vishvas Chalishazar of Pacific Northwest National Laboratories.

\section{Abbreviations}

The following abbreviations are used in this manuscript:\\

\noindent 
\begin{tabular}{@{}ll}
CSZ & Cascadia Subduction Zone \\
PGA & Peak Ground Acceleration \\
BB & Bus Branch power system model \\
xBB & extended Bus Branch power system model \\
SBSB & single bus single breaker \\
RB & ring bus \\
BAH & breaker and a half \\
DBDB & double bus double breaker \\
CB & circuit breaker \\
SD & switch disconnect \\
SG & switch group \\
\end{tabular}

\printbibliography

@article{Frankel:2018:Broadband-synthetic,
	author = {A. Frankel and E. Wirth and N. Marafi and J. Vidale and W. Stephenson},
	journal = {Bulleting of the Seismological Society of America},
	number = {5A},
	pages = {2347-2369},
	title = {Broadband synthetic seismograms for magnitude 9 earthquakes on the Cascadia megathrust based on 3D simulations and stochastic synthetics, part 1: Methodology and overall results},
	volume = {108},
	year = {2018}}

@article{Wirth:2020:Ensemble-ShakeMaps,
	author = {Wirth, Erin A. and Grant, Alex and Marafi, Nasser A. and Frankel, Arthur D.},
	doi = {10.1785/0220200240},
	eprint = {https://pubs.geoscienceworld.org/ssa/srl/article-pdf/92/1/199/5209885/srl-2020240.1.pdf},
	issn = {0895-0695},
	journal = {Seismological Research Letters},
	month = {11},
	number = {1},
	pages = {199-211},
	title = {{Ensemble ShakeMaps for Magnitude 9 Earthquakes on the Cascadia Subduction Zone}},
	url = {https://doi.org/10.1785/0220200240},
	volume = {92},
	year = {2020}}

@article{Heath:2020:A-global-hybrid,
	author = {D.C. Heath and D.J. Wald and C.B. Worden and E.M. Thompson and G.M. Smoczyk},
	journal = {Earthquake Spectra},
	title = {A global hybrid Vs30 map with a topographic slope-based default and regional map insets},
	year = {2020}}

@techreport{Rojahn:1991:Seismic-Vulnerability,
	author = {C. Rojahn},
	institution = {FEMA},
	month = {Sep},
	number = {ATC-25},
	title = {Seismic Vulnerability and Impact of Disruption of Lifelines in the Conterminous United States},
	year = {1991}}

@techreport{ieee-693-2018,
	author = {IEEE Power Engineering Society},
	title = {{IEEE} Standard 693: Recommended Practice for Seismic Design of Substations},
	year = {2018}}

@article{Pires:1995aa,
	author = {J.A. Pires and A.H.S. Ang and R. Villaverde},
	journal = {Nuclear Engineering and Design},
	pages = {427-439},
	title = {Seismic reliability of electrical power transmission systems},
	year = {1995}}

@techreport{Oikonomou:2016aa,
	author = {K. Oikonomou and M. Constantinou and A.M. Reinhorn and L. Kempner},
	institution = {MCEER},
	number = {MCEER-16-0006},
	title = {Seismic isolation of high-voltage electrical power transformers},
	year = {2016}}

@techreport{Huo:1995:Seismic-Fragility,
	author = {J.R. Huo and H.H.M. Hwang},
	institution = {National Center for Earthquake Engineering Research (NCEER)},
	month = {August},
	number = {NCEER-95-0014},
	title = {Seismic Fragility Analysis of Equipment and Structures in a Memphis Electric Substation},
	year = {1995}}

@techreport{Ersoy:2008aa,
	author = {S. Ersoy and B. Feizi and A. Ashrafi and M. Ala Saadeghvaziri},
	institution = {MCEER},
	number = {MCEER-08-0011},
	title = {Seismic evaluation and rehabilitation of critical components of electrical power systems},
	year = {2008}}

@article{S.A.Zareei:2017aa,
	author = {S.A.Zareei and M. Hosseini and M. Ghafory-Ashtiany},
	journal = {Soil Dynamics and Earthquake Engineering},
	pages = {79-94},
	title = {Evaluation of power substation equipment seismic vulnerability by multivariate fragility analysis: a case study on a 420 kV circuit breaker},
	year = {2017}}

@article{Kwasinski:2014aa,
	author = {A. Kwasinski and J. Eidinger and A. Tang and C. Tudo-Bornareld},
	journal = {Earthquake Spectra},
	month = {Feb},
	number = {1},
	pages = {205-230},
	title = {Performance of electric power systems in the 2010-2011 Christchurch, New Zealand, Earthquake Sequence},
	volume = {30},
	year = {2014}}

@article{Park:2006:Nisqually-Earthquake,
	author = {J. Park and N. Nojima and D. Reed},
	journal = {Earthquake Spectra},
	month = {May},
	pages = {491-509},
	title = {Nisqually Earthquake Electric Utility Analysis},
	year = {2006}}

@article{Kazama:2012aa,
	author = {M. Kazama and T. Noda},
	journal = {Soils and Foundation},
	title = {Damage statistics (Summary of the 2011 off the Pacific Coast of Tohoku Earthquake damage)},
	year = {2012}}

@techreport{Asfura:1999:The-Quindio-Columbia,
	author = {A. Asfura},
	institution = {Multidisciplinary Center for Earthquake Engineering Research (MCEER)},
	month = {Oct},
	title = {The Quindio, Columbia Earthquake of January 25, 1999: Reconnaissance Report},
	year = {1999}}

@techreport{Goltz:1994:The-Northridge-California,
	author = {J. Goltz},
	institution = {National Center for Earthquake Engineering Research},
	month = {March},
	number = {NCEER-94-0005},
	title = {The Northridge, California Earthquake of January 17, 1994: General Reconnaissance Report},
	year = {1994}}

@techreport{Schiff:1998:The-Loma-Prieta,
	author = {A. Schiff},
	institution = {US Geological Survey},
	title = {The Loma Prieta, California, Earthquake of October 17, 1989 - Lifelines},
	year = {1998}}

@techreport{Lee:2000:The-Chi-Chi-Taiwan,
	author = {G. Lee},
	institution = {Multidisciplinary Center for Earthquake Engineering Research (MCEER)},
	month = {April},
	number = {MCEER-00-0003},
	title = {The Chi-Chi, Taiwan Earthquake of September 21, 1999: Reconnaissance Report},
	year = {2000}}

@techreport{Cimellaro:2013:Emilia-Earthquake,
	author = {G. Cimellaro and M. Chiriatti and A. Reinhorn and L. Tirca},
	institution = {Multidisciplinary Center for Earthquake Engineering Research (MCEER)},
	month = {June},
	number = {MCEER-13-0006},
	title = {Emilia Earthquake of May 20, 2012 in Northern Italy: Rebuilding A Community Resilient to Multiple Hazards},
	year = {2013}}

@techreport{hazus,
	institution = {FEMA},
	month = {July},
	title = {Hazus 5.1: Hazus Earthquake Model Technical Manual},
	year = {2022}}

@article{Panteli:2017aa,
	author = {M. Panteli and P. Mancarella and D. Trakas and E. Kyriakides and N. Hatziargyriou},
	journal = {IEEE Transactions on Power Systems},
	month = {Nov},
	number = {6},
	pages = {4732-4742},
	title = {Metrics and Quantification of Operational and Infrastructure Resilience in Power Systems},
	volume = {32},
	year = {2017}}

@article{Panteli:2017ab,
	author = {M. Panteli and D. N. Trakas and P. Mancarella and N. Hatziargyriou},
	journal = {Proceedings of the IEEE},
	number = {5},
	pages = {1202-1213},
	title = {Power Systems Resilience Assessment: Hardening and Smart Operational Enhancement Strategies},
	volume = {105},
	year = {2017}}

@techreport{orp,
	author = {K. Yu and et al.},
	institution = {OSSPAC},
	title = {The Oregon Resilience Plan},
	year = {2013}}

@techreport{exercise,
author={},
institution={Washington and Oregon Whole Community Exercise Design Committee},
title={Cascadia Subduction Zone Catastrophic Earthquake and Tsunami},
year={2015}}

@techreport{crew,
author={},
institution={Cascadia Region Earthquake Workgroup},
title={Cascadia Subduction Zone Earthquakes: A Magnitude 9.0 Earthquake Scenario},
year={2013}}

@techreport{goldfinger,
author = {C. Goldfinger and Ann E. Morey and Joel E. Johnson and Jason R. Patton and Eugene B. Karabanov and Julia Gutierrez-Pastor and Andrew T. Eriksson and Eulalia Gracia and Gita Dunhill and Randolph J. Enkin and Audrey Dallimore and Tracy Vallier and Robert Kayen},
institution={U.S. Geological Survey},
title={Turbidite event history -- Methods and implications for Holocene paleoseismicity of the Cascadia subduction zone},
doi = {10.3133/pp1661F}}

@ARTICLE{Birchfield2018,
author={Birchfield, Adam B. and Xu, Ti and Overbye, Thomas J.},  journal={IEEE Transactions on Power Systems},   title={Power Flow Convergence and Reactive Power Planning in the Creation of Large Synthetic Grids},   year={2018},  volume={33},  number={6},  pages={6667-6674},  doi={10.1109/TPWRS.2018.2813525}}

@ARTICLE{Birchfield2017,  
author={Birchfield, Adam B. and Xu, Ti and Gegner, Kathleen M. and Shetye, Komal S. and Overbye, Thomas J.},  journal={IEEE Transactions on Power Systems},   title={Grid Structural Characteristics as Validation Criteria for Synthetic Networks},   year={2017},  volume={32},  number={4},  pages={3258-3265},  doi={10.1109/TPWRS.2016.2616385}}

@inproceedings{chalishazar_augmenting_2018,
	address = {Portland, OR, USA},
	title = {Augmenting the {Traditional} {Bus}-{Branch} {Model} for {Seismic} {Resilience} {Analysis}},
	isbn = {978-1-4799-7312-5},
	url = {https://ieeexplore.ieee.org/document/8557820/},
	doi = {10.1109/ECCE.2018.8557820},
	language = {en},
	urldate = {2022-08-08},
	booktitle = {2018 {IEEE} {Energy} {Conversion} {Congress} and {Exposition} ({ECCE})},
	publisher = {IEEE},
	author = {Chalishazar, Vishvas and Johnson, Brandon and Cotilla-Sanchez, Eduardo and Brekken, Ted K.A.},
	month = sep,
	year = {2018},
	pages = {1133--1137},
}

@ARTICLE{MATPOWER,
author={Zimmerman, Ray Daniel and Murillo-Sánchez, Carlos Edmundo and Thomas, Robert John},  journal={IEEE Transactions on Power Systems},   title={MATPOWER: Steady-State Operations, Planning, and Analysis Tools for Power Systems Research and Education},   year={2011},  volume={26},  number={1},  pages={12-19},  doi={10.1109/TPWRS.2010.2051168}}

@inproceedings{PMAPS_2022,
	address = {Manchester, United Kingdom},
	title = {Seismic {Resilience} {Assessment} of {Electric} {Power} {Systems} {Using} a {Substation} {Bay}-level {Model}},
	isbn = {978-1-66541-211-7},
	url = {https://ieeexplore.ieee.org/document/9810625/},
	doi = {10.1109/PMAPS53380.2022.9810625},
	language = {en},
	urldate = {2022-08-07},
	booktitle = {2022 17th {International} {Conference} on {Probabilistic} {Methods} {Applied} to {Power} {Systems} ({PMAPS})},
	publisher = {IEEE},
	author = {Villamarin-Jacome, Alex and Moreno, Rodrigo},
	month = jun,
	year = {2022},
	pages = {1--6},
}

@article{wang_2016,
	title = {Research on {Resilience} of {Power} {Systems} {Under} {Natural} {Disasters}—{A} {Review}},
	volume = {31},
	issn = {1558-0679},
	doi = {10.1109/TPWRS.2015.2429656},
	number = {2},
	journal = {IEEE Transactions on Power Systems},
	author = {Wang, Yezhou and Chen, Chen and Wang, Jianhui and Baldick, Ross},
	month = mar,
	year = {2016},
	note = {Conference Name: IEEE Transactions on Power Systems},
	pages = {1604--1613},
}

@INPROCEEDINGS{Hines_2008,
author={Hines, Paul and Apt, Jay and Talukdar, Sarosh},  booktitle={2008 IEEE Power and Energy Society General Meeting - Conversion and Delivery of Electrical Energy in the 21st Century},   title={Trends in the history of large blackouts in the United States},   year={2008},  volume={},  number={},  pages={1-8},  doi={10.1109/PES.2008.4596715}}

\if@endfloat\clearpage\processdelayedfloats\clearpage\fi






\end{document}